# Zeno logic gates using micro-cavities

**J.D. Franson, B.C. Jacobs, and T.B. Pittman**

Johns Hopkins University, Applied Physics Laboratory, Laurel, MD 20723

The linear optics approach to quantum computing has several potential advantages but the logic operations are probabilistic.  Here we review the use of the quantum Zeno effect to suppress the intrinsic failure events in these kinds of devices, which would produce deterministic logic operations without the need for ancilla photons or high-efficiency detectors.  The potential advantages of implementing Zeno gates using micro-cavities and electromagnetically-induced transparency is discussed.

## 1. INTRODUCTION

There has been considerable progress in a linear optics approach to quantum computing [1,2], including the experimental demonstration of logic gates [3-6], small-scale circuits [7], quantum error correction [8], and cluster states [9].  As first proposed by Knill, Laflamme, and Milburn [1], probabilistic logic operations can be performed using linear optical elements, ancilla photons, and feed-forward control based on measurements made on the ancilla.  In principle, the intrinsic failure rate of these logic gates can be made arbitrarily small using a large number of ancilla, but that would require a large increase in the required resources as well as high-efficiency detectors.  The use of cluster states [9] appears to be a promising approach, since probabilistic logic operations can be used to set up the initial cluster state, after which an arbitrary quantum calculation can be performed without any intrinsic failure events.

We recently proposed an alternative approach [10] in which the intrinsic failure events of linear optics logic gates can be suppressed using the quantum Zeno effect [11] based on strong two-photon absorption. In addition to providing a brief review of the basic Zeno gate approach, this paper discusses the use of micro-cavities to enhance the two-photon absorption rate and increase the performance of the logic gates; this possibility was briefly mentioned in Ref. [10] without any discussion of how it could be implemented. We will also discuss the use of a new technique for reducing the rate of single-photon scattering that is analogous to electromagnetically-induced transparency, except that it does not require the use of strong laser beams [12].

Since they rely on two-photon absorption, Zeno gates are no longer linear optical devices. They can be viewed as a hybrid approach that combines some aspects of a linear optics approach with the use of a nonlinear medium. This offers some potential advantages over other optical approaches, such as cavity QED [13]. For example, there is no need to convert stationary qubits to flying qubits and back again, and the quantum information remains in the form of single photons at all times. In addition, the necessary two-photon absorption can be achieved using an atomic vapor instead of a single trapped atom or ion, which may simplify the experimental apparatus.

## 2. LINEAR OPTICS LOGIC GATES

Linear optics quantum computing was first discussed by Knill, Laflamme, and Milburn [1] based on a set of nested interferometers. Koashi et al. [14] proposed a similar approach based on polarization-encoded qubits at about the same time, and we introduced a simplified CNOT gate shortly thereafter [15]. A generic linear optics logic gate is illustrated in Fig. 1. Two logical qubits in the form of single photons are mixed with an arbitrary number of ancilla photons in a "black box" containing only linear optical elements, such as beam splitters and phase shifters. The state of the ancilla is measured after they leave the device, and the goal is to design the optical elements in such a way that the quantum measurement process projects out the desired logical output state. This will be the case for some of the measurement results after single-qubit corrections have been applied (feed-forward control), while other measurement outcomes will correspond to output states that are known to be incorrect and cannot be corrected. Incorrect results of that kind will be referred to as failure events, and they occur with a probability that depends on the design of the specific device.

As an example, the CNOT gate that we proposed [15] is shown in Fig. 2. This is a relatively simple and stable device in which the two input qubits are mixed with an entangled pair of ancilla photons using two polarizing beam splitters. The polarization states of the photons that emerge into the two detectors are measured in a rotated basis in order to avoid a determination of their logical value. It can be shown that this device will correctly implement a CNOT logic operation provided that one and only one photon is detected in each detector, which occurs with a probability of ¼.

The CNOT gate of Fig. 2 will fail if two photons emerge from the device in the same optical path, such as two photons entering the top detector. This is the only failure mode of this device, and it would operate properly all of the time if the probability of two photons emerging in the same path could be suppressed in some way.

## 3. ZENO GATES

In the quantum Zeno effect [11], a randomly occurring event can be suppressed by frequent measurements to determine whether or not it has occurred. The basic idea is to use this effect to suppress the emission of two photons into the same output path of our CNOT gate, which would eliminate the intrinsic failure events and produce a deterministic logic gate.

The origin of the Zeno effect is illustrated in Fig. 3. For sufficiently short amounts of time, the probability amplitude for a failure event will increase linearly, as illustrated in Fig. 3a, while the probability itself increases quadratically. If a measurement is made after a short amount of time to determine whether or not an error has occurred, it will be found with high probability that no error has occurred, as illustrated in Fig. 3b. The quantum state of the system will then collapse into the original state with no error, and a sequence of such measurements will continually "reset" the system and prevent an error from occurring.

In order for the Zeno effect to work, the probability amplitude for an error must increase linearly in time, which is not the case for the beam splitters in the CNOT gate of Fig. 2. In order to solve this problem, the beam splitters can be replaced with coupled optical fibers as illustrated in Fig. 4. Here the cores of two optical fibers are brought sufficiently close to each other that their evanescent fields will gradually couple a photon in one fiber core into the other fiber. Devices of this kind are commercially available in the form of 50-50 couplers, for example.

The probability that two photons will be coupled into the same fiber core could be suppressed, at least in principle, by making frequent measurements to determine whether or not that is the case. The dots in Fig. 5 show the results of a calculation [10] in which $N$ such measurements were made during the time that the photons propagated through the coupled set of fibers; it was assumed that the measurement process had no effect if only a single photon were present in each fiber. It can be seen that a sequence of measurements of this kind can suppress the probability that two photons will emerge in the same optical fiber.

As a practical matter, equivalent results can be obtained if atoms that can absorb two photons but not one are placed in the optical fiber cores or in their evanescent fields. Roughly speaking, the atoms "watch" for the presence of two photons in the same fiber core, which should inhibit that from occurring. The solid line in Fig. 5 shows the results of a density matrix calculation in which the probability of a failure event (in which two photons emerge from the same path or one or more photons are absorbed) as a function of the strength of the two-photon absorption rate (in arbitrary units). It can be seen that no actual measurements are required and that strong two-photon absorption is sufficient for the implementation of Zeno logic gates.

Although coupled optical fibers combined with two-photon absorption could be used to implement the CNOT gate of Fig. 2, it can be shown [10] that the coupled-fiber device of Fig. 4 can implement a universal two-qubit logic gate by itself. Here the length of the device is chosen in such a way that a single photon entering one path will be transferred completely to the other path (a SWAP operation), which is accompanied by a phase shift of $\pi/2$. But if two photons are incident, one in each path, the Zeno effect eliminates any coupling between the two fiber cores and no phase shift occurs. As a result, the device of Fig. 4 can implement a SWAP' operation that is equivalent to an ordinary SWAP combined with a controlled phase gate. This is a universal two-qubit operation that can be combined with single-qubit operations to produce a CNOT gate or any other operation.

The main challenge in implementing Zeno gates of this kind will be to achieve a sufficiently large amount of two-photon absorption while keeping the single-photon losses to a small level. The rate of single-photon loss due to scattering by the atoms will be comparable to the two-photon absorption rate if the diameter of the fiber cores is equal to the wavelength of the photons and if the atoms are randomly distributed [10]; that would be unacceptable for logic operations. The single-photon scattering rate can be reduced substantially below that level if the atoms have a nearly uniform density, such as in a crystal structure, since a uniform medium does not scatter light. Other methods for reducing the single-photon scattering rate compared to the two-photon absorption rate are discussed in the next two sections.

## 4. USE OF MICRO-CAVITIES

The rate of two-photon absorption can be enhanced compared to the rate of single-photon scattering by confining the photons to a resonant cavity with a small mode volume. The electric field associated with a single photon is inversely proportional to the square-root of the mode volume and can reach values on the order of $10^4$ volts/meter in typical cavities [10]. The rate of two-photon scattering is proportional to the fourth power of the field while the rate of single-photon scattering is proportional to the second power of the field, so that the ratio of two-photon absorption to single-photon scattering is inversely proportional to the mode volume.

One potential implementation of a Zeno gate using resonant cavities is illustrated in Fig. 6. Here the input qubits $q_1$ and $q_2$ (single photons) are carried in two wave guides that are coupled to two ring resonators $R_1$ and $R_2$. The two resonators are weakly coupled to each other in a manner that is analogous to the coupling between the two fiber cores in Fig. 4. The couplings are adjusted so that a single photon incident in one wave guide will be completely coupled into the other wave guide through the two resonators. In the presence of strong two-photon absorption, this will implement a SWAP' operation as before.

The geometry of Fig. 6 has several advantages over the coupled optical fibers of Fig. 4. First of all, the overall size of the device is reduced, which would be especially important if an atomic vapor is used for the two-photon absorption. In addition, the reduced mode volume enhances the rate of two-photon absorption compared to the single-absorption rate. As was mentioned above, Zeno gates of this kind do not require the trapping of single atoms, as is the case in cavity QED experiments.

The performance of the Zeno gate could be improved if variable couplings were added between the wave guides and the ring resonators, as illustrated in Fig. 7. The idea is to vary the coupling in such a way that an incident single-photon wave packet could be coupled into one of the resonators and stored for some period of time (as limited by the Q of the cavity). This would allow more time for the SWAP' operation to be performed if the rate of two-photon absorption is limited, and it would ensure that both photons were in either $R_1$ or $R_2$ at the same time. The variable coupling could consist of a smaller ring resonator constructed from a material whose index of refraction could be externally controlled, for example. This would allow the resonant frequency of the smaller ring to be adjusted to control the effective coupling between the wave guides and the larger resonators.

## 5. ELECTROMAGNETICALLY-INDUCED TRANSPARENCY

The rate of single-photon scattering can further reduced by using electromagnetically-induced transparency (EIT). In conventional EIT [16,17], the resonant scattering of photons by a two-level atom can be eliminated by applying a strong laser beam tuned to a different transition. The laser beam splits the initial excited atomic state into two dressed states with slightly different frequencies, and the scattering amplitudes from these two states are equal and opposite. This results in a dramatic decrease in the rate of single-photon absorption on the original transition.

One potential problem in using conventional EIT in quantum Zeno gates is the difficulty in separating a single photon from an intense laser beam, even if the photon and laser beam have different frequencies. A similar reduction in the single-photon scattering can be obtained [12], however, without any laser beams if the photons are tuned between two of the resonant modes of an optical cavity, as illustrated in Fig. 8. Here one or more photons are incident in a wave guide which is strongly coupled to a ring resonator. The resonator in turn is coupled to a large number of two-photon absorbing atoms. (The strong coupling between the resonator and the cavity will reduce the Q of the cavity, which allows a significant response even between two of the resonant modes.) Under these conditions, there is destructive interference between the scattering amplitudes from resonator states $l_R$ and $m_R$, since they correspond to equal and opposite detunings. As a result, there should be no single-photon loss due to scattering from the atoms.

The calculated [12] two-photon absorption rate and single-photon scattering rates for this system are shown in Fig. 9. When the incident photons are tuned between the two resonant modes ($\delta = 0$) there is no single-photon scattering, as expected. On the other hand, the two-photon absorption has a strong peak at the same location provided that the upper level of the atomic transition is resonant with the energy of two photons. Numerical estimates for a realistic set of parameters suggest that the single-photon scattering rate can be four or five orders of magnitude less than the two-photon absorption rate under these conditions, which may allow the operation of Zeno gates with relatively low loss.

## 6. SUMMARY AND ACKNOWLEDGEMENTS

Linear optics is a promising approach for quantum computing but the logic operations have an intrinsic failure rate, which can be overcome using cluster states and other techniques. Here we have reviewed an alternative approach [10] in which the quantum Zeno effect is used to suppress the failure events and produce nearly deterministic logic operations. This reduces the overhead in the amount of resources that are required and it also eliminates the need for high efficiency detectors. The Zeno effect can be implemented using strong two-photon absorption to suppress the emission of two photons into the same output port, which is the only failure mode of our CNOT gate.

One of the challenges in implementing Zeno gates is to produce sufficiently strong two-photon absorption with a minimal rate of single-photon scattering. The single-photon scattering can be reduced compared to the two-photon absorption by using a medium with a uniform density, micro-cavities with small mode volume, and a new form of EIT that does not require any strong laser beams [12]. It may be possible to implement high-performance Zeno logic gates using a combination of these techniques.

This work was funded by the Army Research Office and the Disruptive Technology Office under grant # W911NF-05-0397.

**FIGURE CAPTIONS**

Fig. 1. Generic logic gate using linear optical elements. The two input qubits are mixed with a number of ancilla photons using linear optical elements. Corrections to the output may be applied based on measurements made on the ancilla.

Fig. 2. Controlled-NOT logic gate implemented using two polarizing beam splitters, an entangled pair of ancilla, and two detectors. This gate will produce the correct logical output whenever a single photon is found in both detectors, which occurs with a probability of ¼.

Fig. 3. Basic origin of the quantum Zeno effect, in which frequent measurements to determine whether or not an error has occurred will continuously collapse the state vector of the system back into the initial state corresponding to no errors.

Fig. 4. A Zeno logic gate implemented using two coupled optical fibers and strong two-photon absorption due to atoms in the fiber cores or their evanescent fields. The length of the coupled fibers is chosen in such a way that a single photon in either input port will be completely coupled into the other fiber (a SWAP operation). When two photons are input at the same time, the Zeno effect will prevent two photons from being in the same fiber core, which results in a controlled phase gate in addition to the SWAP.

Fig. 5. Reduction in the intrinsic error (failure) rate of a linear optics logic gate due to the quantum Zeno effect. The dots correspond to the error probability in the device of Fig. 4 as a function of the number of measurements performed. The solid line shows the equivalent error probability as a function of the rate of two-photon absorption (arbitrary units).

Fig. 6. Implementation of a Zeno logic gate (SWAP') using two wave guides coupled to two ring resonators $R_1$ and $R_2$, with two-photon absorbing atoms in the evanescent fields of the resonators. This device is equivalent to the coupled optical fibers of Fig. 4, except that the rate of two-photon absorption is enhanced by the small mode volume of the resonators.

Fig. 7. Zeno logic gate implemented as in Fig. 6 but with variable couplings included between the wave guides and the resonators to improve the performance.

Fig. 8. Optical transparency using interference between two modes of a resonant cavity. One or more photons in a wave guide are strongly coupled into a ring resonator, which in turn is coupled into a large number of two-photon absorbing atoms. Quantum interference eliminates single-photon scattering when the incident photons are tuned between two resonator levels, while strong two-photon absorption can still occur.

Fig. 9. Numerical results (solid line) showing a reduction in the single-photon scattering rate due to quantum interference in the device of Fig. 8. The dotted line shows the two-photon absorption rate (both in arbitrary units).

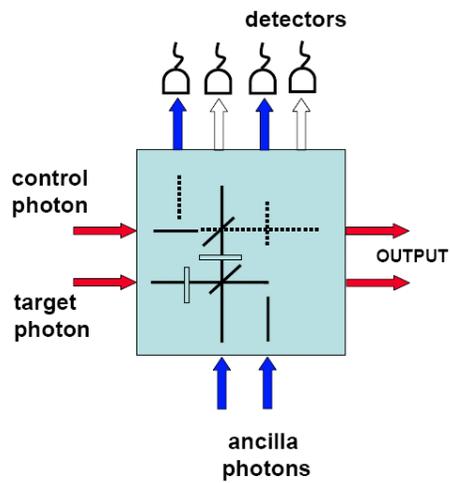

Fig. 1. Generic logic gate using linear optical elements. The two input qubits are mixed with a number of ancilla photons using linear optical elements. Corrections to the output may be applied based on measurements made on the ancilla.

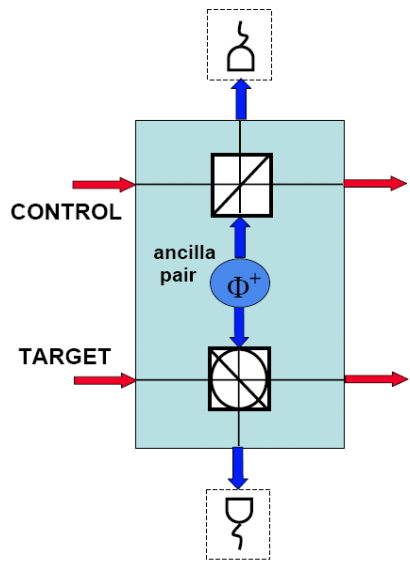

Fig. 2. Controlled-NOT logic gate implemented using two polarizing beam splitters, an entangled pair of ancilla, and two detectors.

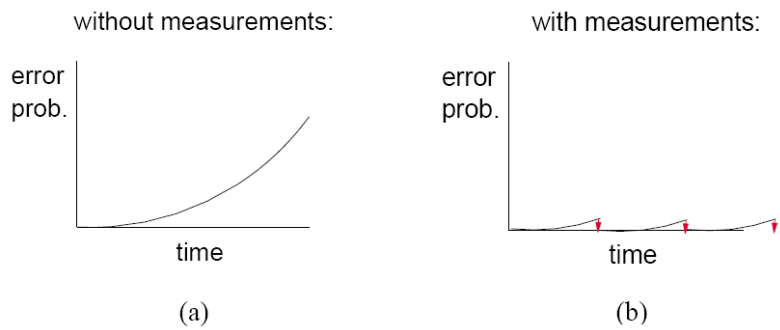

Fig. 3. Basic origin of the quantum Zeno effect, in which frequent measurements to determine whether or not an error has occurred will continuously collapse the state vector of the system back into the initial state corresponding to no errors.

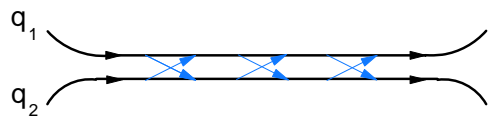

Fig. 4. A Zeno logic gate implemented using two coupled optical fibers and strong two-photon absorption due to atoms in the fiber cores or their evanescent fields. The length of the coupled fibers is chosen in such a way that a single photon in either input port will be completely coupled into the other fiber (a SWAP operation). When two photons are input at the same time, the Zeno effect (due to two-photon absorption) will suppress any interaction and a SWAP plus a controlled phase operation is obtained (SWAP').

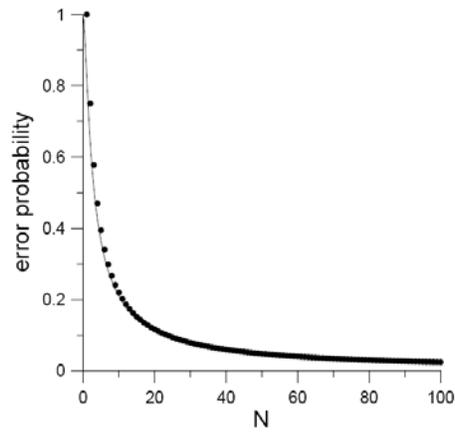

Fig. 5. Reduction in the intrinsic error (failure) rate of a linear optics logic gate due to the quantum Zeno effect. The dots correspond to the error probability in the coupled fiber device of Fig. 4 as a function of the number of measurements performed. The solid line shows the equivalent error probability as a function of the rate of two-photon absorption (arbitrary units).

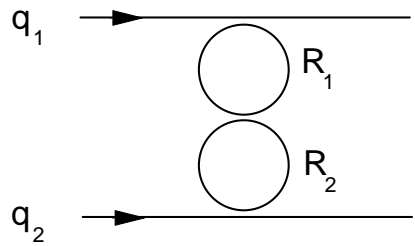

Fig. 6. Implementation of a Zeno logic gate (SWAP') using two wave guides coupled to two ring resonators $R_1$ and $R_2$, with two-photon absorbing atoms in the evanescent fields of the resonators. This device is equivalent to the coupled optical fibers of Fig. 4, except that the rate of two-photon absorption is enhanced by the small mode volume of the resonators.

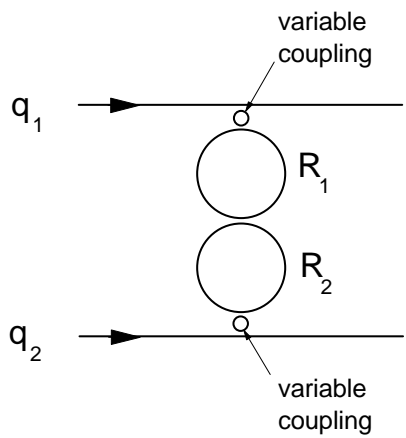

Fig. 7. Zeno logic gate implemented as in Fig. 6 but with variable couplings between the wave guides and the resonators included to improve the performance.

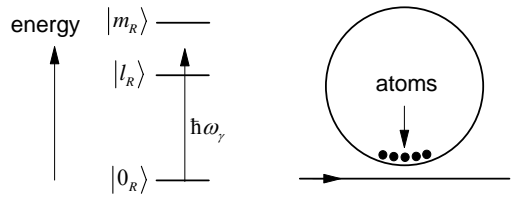

Fig. 8. Optical transparency using interference between two modes of a resonant cavity. One or more photons in a wave guide are strongly coupled into a ring resonator, which in turn is coupled into a large number of two-photon absorbing atoms. Quantum interference eliminates single-photon scattering when the incident photons are tuned between two resonator levels, while strong two-photon absorption can still occur.

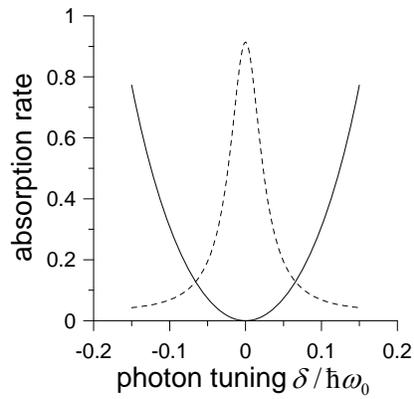

Fig. 9. Numerical results (solid line) showing a reduction in the single-photon scattering rate due to quantum interference in the device of Fig. 8. The dotted line shows the two-photon absorption rate (both in arbitrary units).